\documentclass[3p,twocolumn,authoryear]{elsarticle}
\usepackage[hyphens]{url}
\usepackage{hyperref}

\usepackage{caption}
\usepackage{subcaption}
\usepackage{multirow}
\usepackage{graphicx}
\usepackage{multicol}
\usepackage{array}
\usepackage{comment}
\usepackage{longtable}
\usepackage{amsmath,amssymb,amsfonts}
\usepackage{textcomp}
\usepackage{xcolor}
\usepackage{amsthm}
\usepackage{ulem, soul}
\usepackage{algorithm}
\usepackage{algpseudocode}
\usepackage[capitalize,noabbrev]{cleveref}

\theoremstyle{definition}

\newcolumntype{P}[1]{>{\centering\arraybackslash}p{#1}}

\usepackage{forest}

\begin{document}
\begin{frontmatter}

\title{A Secure Blockchain-Assisted Framework for Real-Time Maritime Environmental Compliance Monitoring}

\author[1]{William C. Quigley}
\ead{wquigley1@fordham.edu}

\author[1]{Mohamed Rahouti\corref{cor1}}
\ead{mrahouti@fordham.edu}

\author[1]{Gary M. Weiss}
\ead{gaweiss@fordham.edu}

\cortext[cor1]{Corresponding author}

\affiliation[1]{organization={Department of Computer and Information Science, Fordham University},
                state={NY}, 
                city={New York},
                country={USA}}

\begin{abstract}
The maritime industry is governed by stringent environmental regulations, most notably the International Convention for the Prevention of Pollution from Ships (MARPOL). Ensuring compliance with these regulations is difficult due to low inspection rates and the risk of data fabrication. To address these issues, this paper proposes a secure blockchain-assisted framework for real-time maritime environmental compliance monitoring. By integrating IoT and shipboard sensors with blockchain technology, the framework ensures immutable and transparent record-keeping of environmental data. Smart contracts automate compliance verification and notify relevant authorities in case of non-compliance. A proof-of-concept case study on sulfur emissions demonstrates the framework's efficacy in enhancing MARPOL enforcement through real-time data integrity and regulatory adherence. The proposed system leverages the Polygon blockchain for scalability and efficiency, providing a robust solution for maritime environmental protection. The evaluation results demonstrate that the proposed blockchain-enhanced compliance monitoring system effectively and securely ensures real-time regulatory adherence with high scalability, efficiency, and cost-effectiveness, leveraging the robust capabilities of the Polygon blockchain.
\end{abstract}

\end{frontmatter}

\section{Introduction}

Blockchain technology has gained significant attention in recent years due to its potential to revolutionize various industries. In the maritime sector, the implementation of blockchain for compliance monitoring could transform the way regulations are enforced in real-time. By leveraging the immutable and transparent nature of blockchain, it becomes possible to create a framework for monitoring and enforcing regulations in maritime operations with unprecedented levels of accuracy and efficiency.

The adverse environmental impact of ocean pollution is profound and far-reaching, and is significantly exacerbated by marine transport. Pollution from ships contributes to the degradation of marine ecosystems, affects biodiversity, and poses a substantial threat to marine life \citep{mobilik2016marine}. Pollutants such as oil, hazardous chemicals, sewage, and garbage can inflict extensive damage on coral reefs, disrupt the food chain, and result in the death of marine species \citep{walker2019environmental}. The effects of routine shipping operations on ports and coastal oceans present considerable risks to marine ecosystems and coastal economies \citep{ng2010environmental}. Moreover, the emission of sulfur oxides and nitrogen oxides from ship exhausts has measurable adverse impacts on human health, contributing to increased mortality rates \citep{kiihamaki2024effects, raut2022impact}. These environmental challenges underscore the urgent need for effective pollution control measures in the maritime industry.

The maritime industry, which is highly international in nature, is governed by the United Nation’s International Maritime Organization (IMO). The IMO consists of 176 member states and 3 associate members and is responsible for setting regulations to ensure the safety, security, and environmental sustainability of maritime operations through a series of conventions \citep{international_maritime_organization_frequently_asked_questions_dbf30dbd}. Member states enforce the provisions of the IMO conventions upon ships registered to the flag of the nation, set the penalties for infringements, and are responsible for imposing penalties for offenses which occur in international waters. However, when transiting through the territorial waters of another state, the port-state has authority to inspect the ships of other nations for compliance with the conventions and take action in accordance with it’s own law \citep{conventions_3b6ab2c6}.

This paper will primarily focus on environmental compliance, particularly the International Convention for the Prevention of Pollution from Ships (MARPOL), which covers the prevention of pollution of the marine environment from operational or accidental causes \citep{752e4617}. Studies and analysis of MARPOL enforcement have identified several shortcomings. For instance, a game-theoretic study \citep{grdovic2022pollute} discussed the pollution decisions of shipowners and countries related to maritime transportation, finding many ships are motivated to pollute due to low inspection rates and expected fines. An analysis of global inspection reports \citep{capt__deepak_mantoju_55c985d7} found that more than 132,000 deficiencies were issued related to MARPOL in the decade from 2009-2019. Additionally, ship's crews may avoid addressing issues until an anticipated inspection. Often, vessels fly Flags of Convenience (FoC) where they register with a particular nation, despite having no connection to that nation, in order to benefit from lower environmental regulations and standards \citep{anthony_van_fossen_1c740484}. While signatories to MARPOL, FoCs often do not stringently enforce the convention, weakening the international goal of limiting shipping environmental damage \citep{kenneth_button__fa804016}. 
Further, MARPOL relies on record keeping by hand, which enables data fabrication to hide MARPOL violations \citep{davey2021problems}. A recent set of regulations, deemed IMO 2020, regulating shipboard sulfur emissions, has presented new compliance challenges \citep{mellqvist2021best, joung2020imo}. These regulations mandate stricter limits on sulfur emissions, necessitating advanced monitoring techniques to ensure compliance. 

Blockchain is defined as a distributed ledger that keeps a permanent and tamper-proof record of transactions using a decentralized network of nodes, where each node confirms and maintains a copy of the ledger \citep{satoshi_nakamoto_90013da0}. Blockchain employs cryptographic techniques to secure the data and can be encrypted to be accessible only to authorized participants with the corresponding private keys \citep{mohamed_rahouti_aa07229c}. A key application of modern blockchain is smart contracts, which are self-executing contracts that autonomously execute and enforce stipulated rules without intermediaries \citep{anisha_miah_86460258, shi_yi_lin_08549185}.

Ships are equipped with several systems for monitoring the discharge of engine exhaust gases, oily waste, ballast water, sewage, and other sources for the purpose of ensuring compliance with MARPOL. Combining data from IoT and other electronic shipboard sensors with transmission to the blockchain, smart contracts can determine the real-time compliance status of these systems and notify the flag state and the port state in which non-compliance occurs.  Blockchain technology is well-suited for maritime environmental compliance monitoring due to its decentralized and transparent nature, which allows all relevant parties to access the same real-time data. This facilitates efficient monitoring and enforcement of regulations, enhancing environmental protection. The immutability of blockchain records ensures data integrity and trust in compliance monitoring. Additionally, smart contracts enable automated enforcement of regulations, reducing the potential for human error and ensuring swift response to non-compliance issues. Hence, the proposed blockchain can both mitigate data fabrication and promptly notify port-states allowing local regulatory agencies to enforce penalties, solving two of the major issues in MARPOL enforcement.

\begin{figure*}[h!]
  \centering
  \includegraphics[width=\linewidth]{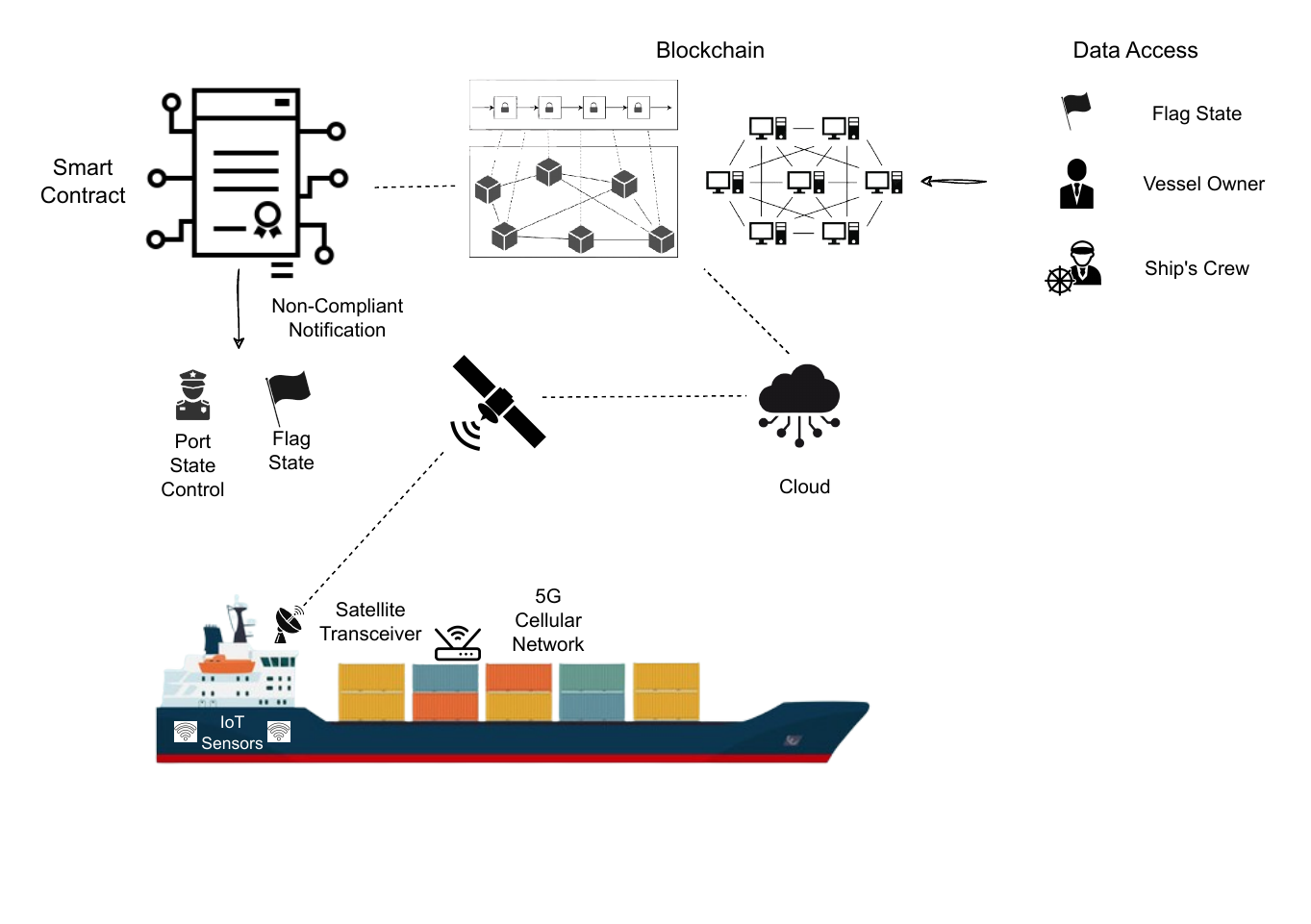} 
  \caption{Conceptual framework for blockchain-enabled maritime environmental compliance monitoring.}
  \label{fig:conceptl}
\end{figure*}

In this work, we propose a conceptual framework for environmental maritime compliance monitoring, leveraging sensor data, blockchain technology, and smart contracts as illustrated in Fig. \ref{fig:conceptl}. Additionally, we present a proof-of-concept case study focusing on sulfur emissions, which validates the theoretical soundness of the proposed framework in a laboratory setting. This system prototype is implemented using Polygon Proof-of-Stake (PoS), a protocol and framework designed for building and connecting Ethereum-compatible blockchain networks, offering enhanced scalability and security through its sidechain and Layer 2 scaling solutions \citep{polygon}. Our solution reduces the likelihood of data fabrication and automates the reporting of non-compliant activities, significantly improving MARPOL enforcement.

The key contributions of this paper are outlined as follows:
\begin{itemize}
    \item Propose a conceptual blockchain-assisted framework for maritime environmental compliance monitoring using data collection via shipboard sensors, and smart contracts that automate compliance checking. 
    \item Develop a prototype using Polygon PoS and verify proof of concept through the application of sulfur emission monitoring.
    \item Explore the practicality of the architecture through a performance assessment and security evaluation.
\end{itemize}

The remainder of this paper is organized as follows. Section \ref{sec:related} provides a comprehensive overview of previous research and recent advancements in the field. Section \ref{sec:method} explores the conceptual architectural design and methodology in detail. Section \ref{sec:evaluation} describes the proof-of-concept experiment design and presents the performance results. Subsequently, Section \ref{sec:discussion} highlights our key findings, discusses the limitations, suggests directions for future research and implications of this work.

\section{Background and Related Work} \label{sec:related}
\subsection{Blockchain Technology}

Blockchain technology, fundamentally a decentralized digital ledger, securely records data across a distributed network of nodes, ensuring immutability and transparency. Data on the blockchain is secured using cryptographic techniques and can be encrypted to be accessible only to authorized participants possessing the corresponding private keys. Critical to the application of our model are the concepts of public blockchains, consensus mechanisms, and smart contracts.

The primary classifications of blockchain applications are public and private blockchains \citep{mohan2019state}. Public blockchains, exemplified by Bitcoin and Ethereum, are open to anyone who wishes to join and participate in the network \citep{ali2019blockchain}. These blockchains are distinguished by their decentralization and transparency, allowing any participant to verify transactions and maintain the ledger. Public blockchains commonly utilize consensus mechanisms such as Proof of Work (PoW) and Proof of Stake (PoS). PoW requires miners to solve complex cryptographic puzzles to validate transactions, which is resource-intensive. In contrast, PoS selects validators based on the number of tokens they are willing to stake, providing a more energy-efficient alternative to PoW \citep{lashkari2021comprehensive}.

Smart contracts are another vital component of blockchain enabled applications. These self-executing contracts have the terms of the agreement directly written into code, automatically enforcing and executing the contract when predefined conditions are met. Smart contracts eliminate the need for intermediaries, reduce transaction costs, and enhance efficiency. They are integral to numerous blockchain applications, facilitating automated, transparent, and tamper-proof processes \citep{shi_yi_lin_08549185}.

For the application of blockchain in maritime environmental compliance, we propose utilizing a permissioned blockchain built on Polygon PoS with smart contracts. Polygon PoS provides a scalable and efficient consensus mechanism, which ensures high-throughput operations. By using smart contracts, Polygon PoS can enforce permissions and manage data replication across the network securely. Polygon’s framework supports robust security features and scalability, making it an effective solution for sensor based blockchain applications. Polygon PoS is an efficient solution for IoT blockchain applications, as demonstrated by Rana et al. \citep{rana2023decentralized}, who used smart contracts on the Polygon blockchain to address issues like data alteration and unauthorized access, ensuring data integrity and security of digital evidence in the judicial system.

This model employs a predefined ledger of trusted devices, eliminating the need for resource-intensive PoW algorithms and reducing power consumption. Furthermore, this approach aligns with the IMO and member-state model, where member states and IMO registered ships act as trusted, pre-approved validators, and ships with registered IMO numbers are have approved machinery installations with required sensors. This framework leverages the strengths of blockchain technology to enhance real-time compliance monitoring and regulatory enforcement in maritime operations.

\subsection{Blockchain for Environmental Monitoring}

Blockchain technology has been demonstrated to be effective in applications for environmental monitoring. Yan et al. \citep{yan2019environmental} presented an environmental monitoring system utilizing biotechnology sensing technology and blockchain to address issues with traditional environmental monitoring, such as data fraud, interference, and inconsistency in environmental data. 

Allena \citep{ allena2020blockchain} examined the impact of blockchain technology on monitoring compliance with environmental regulations, highlighting its potential to enhance data management, transparency, and efficiency by enabling dispersed checks and involving various non-public actors in the process. The study argues that blockchain can significantly improve the verification of environmental data and support more effective enforcement of environmental laws.

Liao et al. \citep{liao2021securing} proposed a blockchain-enabled framework for collaborative environmental monitoring in smart cities using a software-defined Internet of Drones. The framework ensures secure and efficient cooperation among drone controllers from different service providers through a consortium blockchain. A case study demonstrated the effectiveness of the framework in providing trustworthy environmental monitoring and secure data exchange in smart city environments.

Zhong et al. \citep{ zhong2022blockchain} proposed a blockchain-enabled framework for on-site construction environmental monitoring to address limitations in centralized systems, such as information imbalance and data disputes. The framework collects pollutant data via sensors and uploads it to a blockchain network, where smart contracts autonomously monitor pollution levels and evaluate performance. A case study using Hyperledger Fabric validates the framework's viability, demonstrating blockchain's potential to provide trustworthy environmental data and continuous monitoring in real-world scenarios.

\subsection{Maritime Blockchain Applications}

The integration of blockchain technology into maritime operations has garnered significant attention in recent years \citep{lasmoles2021impacts}. Several studies have explored the potential of blockchain to enhance maritime data management and security. For instance, the AISChain project \citep{duan2022aischain} demonstrated the use of blockchain to secure and verify ship Automatic Identification System (AIS) data, which is crucial for maritime safety and regulatory compliance. Wu et al. \citep{wu2023blockchain} proposed a blockchain-based cryptographic anti-spoofing scheme for GNSS civil navigation messages that effectively resisted spoofing attacks while maintaining low computational costs. Freire et al. \citep{freire2022towards} developed a blockchain-based maritime monitoring system integrating low-cost IoT devices and AIS data.

Blockchain has been proposed to enhance supply chain management objectives such as cost, quality, speed and risk management within ports and vessels \citep{li2021survey, farah2024survey}. Furthermore, the application of smart contracts in maritime operations has shown promise in automating regulatory compliance processes \citep{hasan2019smart}. Smart contracts can automatically execute and verify compliance-related actions based on predefined rules, thus reducing the reliance on manual inspections and minimizing the risk of human error. This approach has been successfully implemented in various pilot projects, demonstrating improved efficiency and accuracy in compliance monitoring \citep{liu2022verifying, wu2024compliance, liu2023blockchain}.

In the context of environmental regulation, blockchain technology has been proposed as a solution to improve the monitoring and enforcement of maritime pollution standards. Czachorowski et al. \citep{czachorowski2019application} explored the potential of blockchain technology to improve environmental efficiency in the maritime industry by reducing pollution and operational costs. They highlighted how blockchain can connect the supply chain more efficiently, provide transparent and tamper-proof data exchange, and enhance compliance with environmental regulations. The study emphasizes the broad applicability of blockchain in facilitating inspections, audits, and overall maritime operations.

Despite these advancements, several challenges remain in the widespread adoption of blockchain for maritime compliance monitoring. Issues such as data interoperability, scalability of blockchain solutions, and the integration of legacy systems with new technologies need to be addressed to fully realize the benefits of blockchain in this sector. Additionally, the regulatory framework for blockchain implementation in maritime operations is still evolving, requiring ongoing collaboration between industry stakeholders and regulatory bodies \citep{farah2024survey, pu2021blockchain}.

\subsection{State-of-the-Art Approaches for Sulfur Emission Enforcement}

Reduction in greenhouse gas emissions is a long-term goal of the IMO \citep{joung2020imo}. Updated by IMO 2020, MARPOL Annex VI, Regulation 14, regulates sulfur emissions from ships. Within an Emissions Control Area (ECA), the permitted sulfur content is no more than 0.10\%; outside of an ECA, the permitted sulfur content is no more than 0.50\%. ECAs include specific latitudes and longitudes within the Baltic Sea Area, the North Sea Area, the North American area, and the United States Caribbean Sea. Ships may meet this requirement by using fuel with low sulfur content or by using an approved exhaust gas cleaning system \citep{752e4617}.

Liu \citep{liu2022supervision} emphasizes the challenges in enforcement and the need for coordinated global efforts to better manage compliance. Knudsen \citep{knudsen2021shipping} found that while IMO 2020 compliance is trending positively, there is still substantial room for improvement. A game-theoretic study by Zis \citep{zis2021game} modeling IMO sulfur cap enforcement suggested higher penalties and mandatory retrofitting for repeated violations to increase compliance. Mellqvist and Jacobo \citep{mellqvist2021best} analyzed remote compliance monitoring techniques for IMO sulfur regulations, highlighting the effectiveness of sniffer systems and optical sensors in monitoring fuel sulfur content without on-board inspections. These methods have significantly reduced non-compliance rates in northern Europe from 5-13\% in 2015 to below 1\% in 2020. Brun and Freiholtz \citep{brun2020methods} provided further evidence promoting the effectiveness of remote sensing.

Our proposed system, utilizing onboard sulfur sensors, is feasible. Wijaya et al. \citep{wijaya2014real} presented a monitoring system comprised of gas analyzers, sensors, a microcontroller, and an Iridium satellite communication module for measuring shipboard greenhouse gas emissions. Further, shipboard sulfur emission sensors are both well established in the literature and continue to be refined \citep{el2018photoacoustic,du2022self,bolbot2020novel}.

Overall, the existing literature highlights the significant potential of blockchain technology to transform maritime regulatory compliance. By leveraging the strengths of blockchain and sensor networks, this research aims to develop a comprehensive framework for real-time compliance monitoring in maritime operations, addressing the current gaps and challenges identified in previous studies. We are not aware of any existing studies that proposed the use of sensors on environmental compliance shipboard systems and blockchain-enabled monitoring of such systems for compliance.

Unlike existing studies, our work aims to propose a blockchain-assisted framework for maritime environmental compliance using sensor data collection and smart contracts with an operational proof of concept prototype evaluated for scalability and security.

\section{Methodology and Framework Design}  \label{sec:method}

Our proposed framework integrates sensor data collection, validation, and blockchain registration to provide real-time monitoring of maritime environmental compliance. Algorithm \ref{alg:maritime_compliance} outlines the key steps, beginning with data retrieval from onboard sensors and proceeding through validation, blockchain registration, and compliance checks.

Initially, each raw sensor reading \(d_i \in D\) undergoes integrity checks (\textit{Validate Data}) to remove invalid or malformed entries. Valid data points are then assigned a unique compliance identifier, timestamped, hashed, and committed to the blockchain (\textit{Register Compliance Data on Blockchain}). The immutable ledger preserves a transparent audit trail of both compliant and non-compliant events. Compliance is assessed through a function \(\text{IS\_COMPLIANT}(d_i)\), with violations triggering automated notifications to relevant stakeholders. The final output is a mapping \(S=\{(d_i, s_i)\}\) that denotes each data point’s compliance status.

In parallel, our mathematical modeling formalizes data integrity and anomaly detection. We represent sensor inputs as tuples \(d_{k,i}\) across discrete time indices \(\mathcal{T}\), and employ one-way hash functions to ensure tamper resistance. Cross-sensor consistency checks identify malfunctions or tampering before data reaches the blockchain, while a Proof-of-Stake consensus mechanism secures the distributed ledger. By combining these elements, the framework offers an end-to-end solution for secure, real-time compliance monitoring in maritime operations.

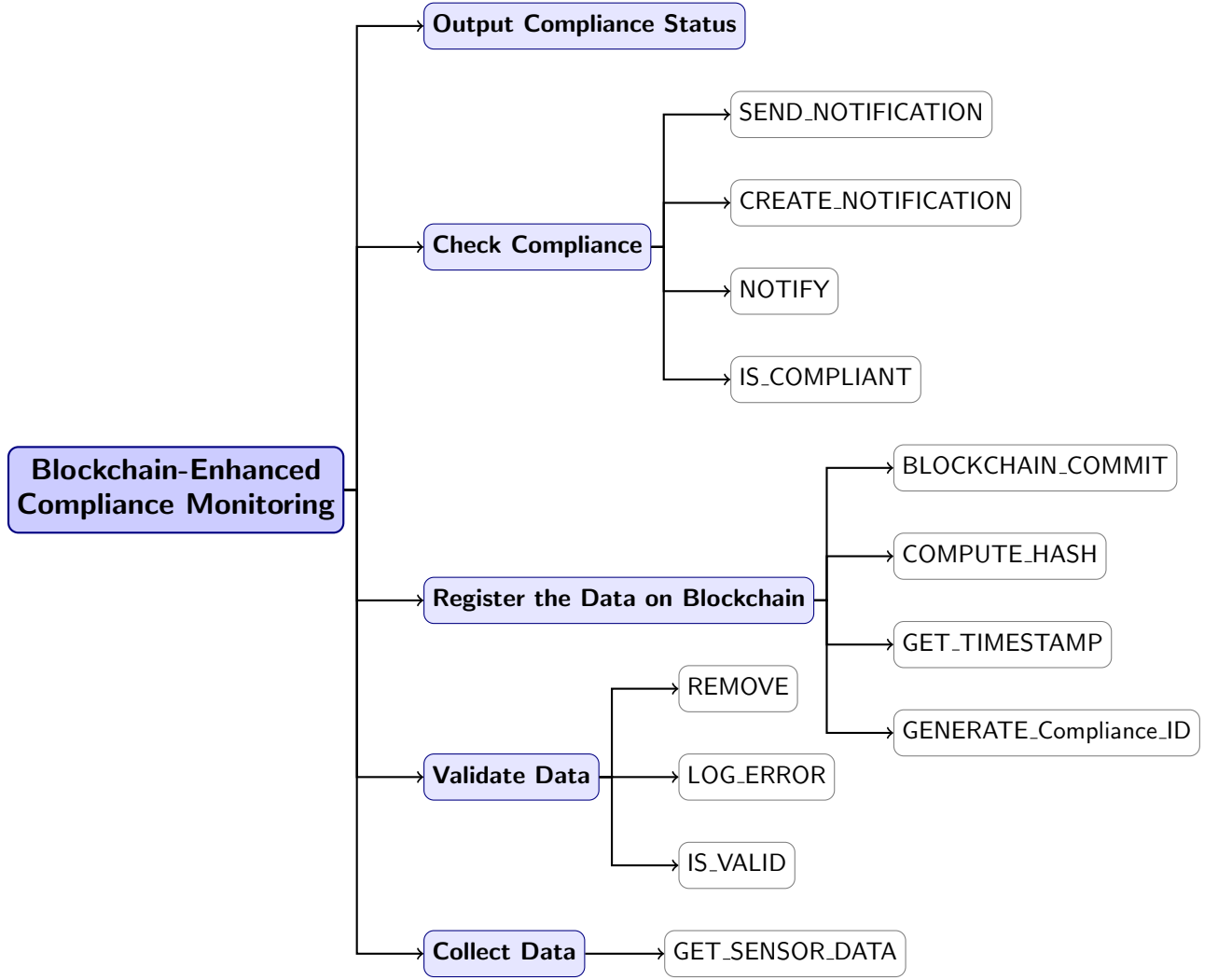
\begin{figure*}[ht]
\centering
\begin{forest}
for tree={
    edge+={thick, ->, black},
    grow=east,
    anchor=west,
    draw,
    rounded corners,
    align=center,
    font=\sffamily,
    parent anchor=east,
    child anchor=west,
    l sep+=20pt,
    s sep+=10pt,
    edge path={
        \noexpand\path [\forestoption{edge}] (!u.parent anchor)
        -- +(5pt,0) |- (.child anchor)\forestoption{edge label};
    }
},
before typesetting nodes={
    if n=1{
        insert before={[,phantom]}, 
    }{}
},
[Blockchain-Enhanced\\Compliance Monitoring,
    fill=blue!20,
    font=\sffamily\bfseries\large,
    draw=blue!50!black,
    thick,
    edge+={thick, -, blue},
    tier=word,
    [Collect Data,
        fill=blue!10,
        draw=blue!50!black,
        font=\sffamily\bfseries,
        [GET\_SENSOR\_DATA,
            fill=white,
            draw=gray
        ]
    ]
    [Validate Data,
        fill=blue!10,
        draw=blue!50!black,
        font=\sffamily\bfseries,
        [IS\_VALID,
            fill=white,
            draw=gray
        ]
        [LOG\_ERROR,
            fill=white,
            draw=gray
        ]
        [REMOVE,
            fill=white,
            draw=gray
        ]
    ]
    [Register the Data on Blockchain,
        fill=blue!10,
        draw=blue!50!black,
        font=\sffamily\bfseries,
        [GENERATE\_Compliance\_ID,
            fill=white,
            draw=gray
        ]
        [GET\_TIMESTAMP,
            fill=white,
            draw=gray
        ]
        [COMPUTE\_HASH,
            fill=white,
            draw=gray
        ]
        [BLOCKCHAIN\_COMMIT,
            fill=white,
            draw=gray
        ]
    ]
    [Check Compliance,
        fill=blue!10,
        draw=blue!50!black,
        font=\sffamily\bfseries,
        [IS\_COMPLIANT,
            fill=white,
            draw=gray
        ]
        [NOTIFY,
            fill=white,
            draw=gray
        ]
        [CREATE\_NOTIFICATION,
            fill=white,
            draw=gray
        ]
        [SEND\_NOTIFICATION,
            fill=white,
            draw=gray
        ]
    ]
    [Output Compliance Status,
        fill=blue!10,
        draw=blue!50!black,
        font=\sffamily\bfseries
    ]
]
\end{forest}
\caption{Key procedures for collecting, validating, registering, and checking compliance data using sensors and blockchain technology.}
\label{fig_taxonomy}
\end{figure*}

\begin{algorithm*}
\caption{Blockchain-enhanced compliance monitoring in maritime operations.}
\label{alg:maritime_compliance}
\begin{algorithmic}[1]

\State \textbf{Input}: Sensor Data $D = \{d_1, d_2, \ldots, d_n\}$ from ship systems
\Comment{$D$: Set of sensor data}
\State \textbf{Output}: A mapping $S = \{(d_i, s_i)\}$ for each monitored activity
\Comment{$S$: Mapping from data point $d_i$ to compliance status $s_i$}

\Procedure{Collect Data}{}
    \For{each $d_i \in D$}
        \State $d_i \gets \text{GET\_SENSOR\_DATA}(d_i)$
        \Comment{Retrieve raw data for $d_i$ from its corresponding sensor.}
    \EndFor
\EndProcedure

\Procedure{Validate Data}{}
    \For{each $d_i \in D$}
        \If{$\neg \text{IS\_VALID}(d_i)$}
            \State $\text{LOG\_ERROR}(\text{'Invalid data: '} + d_i)$
            \Comment{Log an error for invalid data.}
            \State $\text{REMOVE}(d_i \text{ from } D)$
            \Comment{Remove invalid data from $D$.}
        \EndIf
    \EndFor
\EndProcedure

\Procedure{Register Compliance Data on Blockchain}{}
    \For{each $d_i \in D$}
        \State $complianceID \gets \text{GENERATE\_UNIQUE\_ID}()$
        \Comment{Generate a unique compliance ID.}
        \State $d_i.\text{timestamp} \gets \text{GET\_CURRENT\_TIMESTAMP}()$
        \Comment{Assign a current timestamp to $d_i$.}
        \State $d_i.\text{hash} \gets \text{COMPUTE\_HASH}(d_i)$
        \Comment{Compute hash of $d_i$.}
        \State $\text{BLOCKCHAIN\_COMMIT}(d_i, complianceID)$
        \Comment{Store $d_i$ on the blockchain with the generated ID.}
    \EndFor
\EndProcedure

\Procedure{Check Compliance}{}
    \State $S \gets \emptyset$
    \Comment{Initialize the mapping from data points to statuses.}
    \For{each $d_i \in D$}
        \State $c_i \gets \text{IS\_COMPLIANT}(d_i)$
        \Comment{Determine compliance (true/false or \{1,0\}).}
        \If{$c_i = 1$}
            \State $S \gets S \cup \{(d_i, \text{compliant})\}$
        \Else
            \State $S \gets S \cup \{(d_i, \text{non-compliant})\}$
            \State $\text{NOTIFY}(\text{GET\_COMPLIANCE\_ID}(d_i))$
            \Comment{Notify relevant parties using $d_i$'s compliance ID.}
        \EndIf
    \EndFor
    \State \textbf{return} $S$
    \Comment{Return the mapping of each data point to its compliance status.}
\EndProcedure

\Procedure{Notify}{\textit{complianceID}}
    \State $notification \gets \text{CREATE\_NOTIFICATION}(\textit{complianceID})$
    \Comment{Create a notification for the given compliance ID.}
    \State $\text{SEND\_NOTIFICATION}(notification)$
    \Comment{Send the notification to relevant authorities.}
\EndProcedure

\end{algorithmic}
\end{algorithm*}

\subsection{System Modeling}

\subsubsection{Data Points and Hashing}

Consider a discrete set of time indices
\[
\mathcal{T} \;=\; \{ t_1,\, t_2,\, \dots\},
\]
where measurements are taken at each \(t_k \in \mathcal{T}\). Let
\[
\mathcal{D} \;=\; \bigl\{\, d_{k,i} \;\mid\; k \in \mathbb{N},\; 1 \leq i \leq n_k \bigr\}
\]
be the global set of data points, with \(d_{k,i}\) denoting the \(i\)-th data point collected at time \(t_k\). Each data point \(d_{k,i}\) is a tuple:
\[
d_{k,i} \;=\; \bigl(\text{vessel},\ \text{regulation},\ \text{status},\ \text{timestamp}\bigr),
\]
where: \textit{vessel} is a unique identifier (e.g., the IMO number), \textit{regulation} references the applicable MARPOL Annex or environmental requirement defined, \textit{status} is a raw or preliminary indicator of compliance as reported by onboard sensors, and \textit{timestamp} is \(t_k\), the exact time of measurement.

To preserve data integrity, define a one-way cryptographic hash function \(H(\cdot)\). For each \(d_{k,i}\), we compute:
\begin{align*}
h_{k,i} \;=\; H\bigl(d_{k,i}\bigr)
&=\; H\Bigl(
   \text{vessel} \;\parallel\;
   \text{regulation} \;\parallel \\
&\qquad \text{status} \;\parallel\;
   \text{timestamp}
\Bigr),
\end{align*}
where \(\parallel\) denotes concatenation. The collision resistance of \(H\) guarantees that tampering with \(d_{k,i}\) would alter \(h_{k,i}\), making unauthorized modifications detectable.

\subsubsection{Validation and Cross-Sensor Consistency}

For each \(t_k\), we gather a collection of raw data points:
\[
D_{t_k} \;=\; \bigl\{\, d_{k,1},\ d_{k,2},\ \dots,\ d_{k,n_k} \bigr\}.
\]
We apply a validation operator \(\mathrm{VALIDATE}\) to each \(d_{k,i}\), resulting in
\[
V\bigl(D_{t_k}\bigr)
=\left\{
\begin{aligned}
v_{k,i} \;&\mid\; v_{k,i} = \mathrm{VALIDATE}\bigl(d_{k,i}\bigr), \\
&\forall\, d_{k,i} \in D_{t_k}
\end{aligned}
\right\}.
\]

A data point \(v_{k,i}\) is considered valid if it satisfies certain checks, including correct formatting, sensor calibration, and credible numeric ranges (e.g., a sulfur content not exceeding physical possibilities). 

Additionally, to detect sensor malfunction or tampering, let \(\mathcal{S} = \{S_1,\, S_2, \dots\}\) be the set of sensors measuring overlapping parameters. For each \(S_a, S_b \in \mathcal{S}\), define
\[
R_a(t_k),\; R_b(t_k)
\]
as the sensor readings at time \(t_k\). We impose a consistency threshold \(\epsilon > 0\) such that
\[
C_{a,b}(t_k) \;=\;
\begin{cases}
1, &\!\!\text{if } \bigl| R_a(t_k) - R_b(t_k) \bigr| \;\le\; \epsilon,\\
0, &\!\!\text{otherwise}.
\end{cases}
\]
An aggregate score
\[
\bar{C}_{a,b}(\mathcal{T}_w) \;=\;
\frac{1}{|\mathcal{T}_w|}
\sum_{t \in \mathcal{T}_w}
C_{a,b}(t)
\]
over a sliding window \(\mathcal{T}_w \subseteq \mathcal{T}\) helps reveal persistent sensor discrepancies. If \(\bar{C}_{a,b}(\mathcal{T}_w)\) is below a specified threshold \(\gamma\), we mark readings from sensors \(S_a\) or \(S_b\) as suspect and exclude them from further processing.

\subsubsection{Compliance Function}

Define a function
\[
C : \mathcal{D} \;\to\; \{0,1\},
\]
which evaluates whether a validated data point \(d_{k,i}\) meets regulatory standards. For example,
\[
c_{k,i} \;=\; C\bigl(d_{k,i}\bigr),
\]
where
\[
c_{k,i}
\;=\;
\begin{cases}
1, & \text{if } d_{k,i} \text{ complies with the regulation},\\[4pt]
0, & \text{otherwise}.
\end{cases}
\]
\(C\) can incorporate additional domain-specific logic, e.g., verifying that sulfur emission levels remain below 0.50\% outside an Emission Control Area and below 0.10\% inside it.

\subsubsection{Proof-of-Stake Consensus}

We adopt a Proof-of-Stake (PoS) protocol to determine which validator appends new blocks to the ledger. At time \(t_k\), let \(S_i\) be the set of active validators, each holding a stake \(p_j\). The probability \(P_j\) of selecting validator \(j\) to propose the next block is
\[
P_j \;=\;
\frac{\,p_j\,}{\sum_{v \in S_i} p_v}.
\]
Once a validator is chosen, it bundles hashed data points \(\{h_{k,i}\}\) into a candidate block. Other validators verify the block; upon approval by a supermajority, the block is committed to the chain. If the validator acts maliciously (e.g., includes fraudulent data or double-signs), its stake \(p_j\) may be slashed, deterring dishonesty.

\subsubsection{Blockchain Commit Procedure}

At each \(t_k\), after validation and compliance evaluation, the system commits all validated data points to the blockchain along with their compliance status. Define
\[
\Omega_{t_k} \;=\; \Bigl\{\, (h_{k,i},\, c_{k,i}) \;\mid\; d_{k,i}\in V(D_{t_k}) \Bigr\},
\]
where \(h_{k,i}=H(d_{k,i})\) is the cryptographic hash of data point \(d_{k,i}\) and 
\[
c_{k,i} = \begin{cases} 
1, & \text{if } d_{k,i} \text{ is compliant},\\[2mm]
0, & \text{if } d_{k,i} \text{ is non-compliant}.
\end{cases}
\]
Upon majority consensus, the block \(B_{t_k}(\Omega_{t_k})\) is appended to the blockchain, preserving both compliant and non-compliant records. Non-compliant entries are flagged for follow-up actions.

\subsection{Architectural Design and Components}

The architecture of the proposed blockchain-enhanced compliance monitoring system in maritime operations integrates sensors and blockchain technology to ensure data integrity, transparency, and regulatory compliance. The system is designed to address the challenges of data validation, secure storage, and efficient compliance verification in a decentralized environment. Fig. \ref{fig_taxonomy} illustrates the key procedures involved in collecting, validating, registering, and checking compliance data.

\subsubsection{Data collection}

The process begins with the deployment of sensors across various points on the vessel. These sensors continuously monitor and collect data related to ship operations, environmental conditions, and regulatory parameters. At each sampling interval \(t_k\), sensors across different points on the vessel gather raw measurements, forming a set
\(D_{t_k} \;=\; \bigl\{\,d_{k,1},\,d_{k,2},\,\dots,\,d_{k,n_k}\bigr\},\)
where \(d_{k,i}\) corresponds to the \(i\)-th data point collected at time \(t_k\). These data points capture operational, environmental, and regulatory parameters (e.g., emissions or water quality). The system aggregates \(D_{t_k}\) for subsequent validation.

\subsubsection{Sensor Network and Communication Infrastructure}
The proposed framework relies on a robust sensor network deployed across the vessel to continuously monitor operational and environmental parameters, such as sulfur emissions. Data is gathered in real time by sensors connected via onboard wired and wireless protocols to a central processing unit. Within the vessel, an automation and safety network based on industrial control system (ICS) protocols such as Modbus and OPC Unified Architecture (OPC UA) and interfacing with IEC 61162-460-compliant components ensures seamless integration between sensors and actuators essential for monitoring and controlling critical systems \citep{cho2022cybersecurity}.

Sensor data is initially collected over a secure shipboard network employing protocols like TLS for end-to-end encryption. Robust security measures, including encryption of data in transit, secure key management, and integrity checks, are maintained at every communication stage as data is routed through onboard gateways. From these gateways, the data is transmitted to central servers via the ship-to-shore network, which relies on satellite communications using systems such as Inmarsat, Iridium, and VSAT \citep{cho2022cybersecurity}. Moreover, emerging Low Earth Orbit (LEO) mega constellations and High-Throughput Satellites (HTS) are enhancing maritime connectivity by offering higher data rates and increased resilience under adverse conditions \citep{wei2021hybrid, hoyhtya2020integrated}.

\subsubsection{Data validation}

Each raw data point \(d_{k,i} \in D_{t_k}\) is validated. For instance, checks may include proper formatting, plausible numeric ranges, and sensor calibration status. Additionally, cross-sensor consistency ensures that readings from overlapping sensors remain within a predefined threshold. Any invalid data is flagged or removed. The result is a set of validated data points, \( V(D_{t_k}) \;=\; \{\,v_{k,1},\,v_{k,2},\,\dots\},\) which proceeds to the next stage.

\subsubsection{Hash computation and blockchain registration}

Each validated data point \(v_{k,i} \in V(D_{t_k})\) is assigned a cryptographic hash \(h_{k,i} = H(v_{k,i})\) to ensure immutability and tamper detection. The system commits \((v_{k,i},\,h_{k,i})\) to the blockchain. Since the ledger is decentralized, no single party can retroactively alter or remove a data record without being detected by other validators.

\subsubsection{Compliance checking}

Once committed to the ledger, each data point \(v_{k,i}\) undergoes a compliance check using the function \(C\bigl(v_{k,i}\bigr)\), yielding a binary status \((c_{k,i} \;=\; 0, 1)\) Data deemed non-compliant (\(c_{k,i} = 0\)) triggers automated notifications to relevant stakeholders for further investigation.

\subsubsection{Consensus}

Finally, the system leverages a Proof-of-Stake (PoS) consensus mechanism to validate and append new blocks. A validator is selected with probability proportional to its stake to propose a block containing hashes \(\{h_{k,i}\}\). Upon majority approval by other validators, the block is permanently recorded, establishing an auditable trail of both compliant and non-compliant data. 

\subsubsection{Algorithm implementation}

Algorithm \ref{alg:maritime_compliance} outlines the detailed steps involved in the entire process, from data collection to compliance checking. The algorithm ensures that all procedures are systematically executed, maintaining the integrity and accuracy of the compliance monitoring system. Collectively, these steps produce a robust, near real-time compliance monitoring framework suitable for maritime operations.

\subsection{Implementation}
The described blockchain-enhanced compliance monitoring system is implemented on the Polygon PoS network through the use of smart contracts. These smart contracts facilitate the automation of compliance verification processes, ensuring data integrity and transparency. By leveraging the Polygon PoS network, the system benefits from a scalable and efficient platform suitable for the high-throughput requirements of maritime operations.

\subsubsection{Smart contracts for compliance checking}
Smart contracts are central to the compliance monitoring framework, automating the verification and recording of compliance data. Algorithm \ref{alg:sulfur} provides the pseudocode for the sulfur emissions smart contract.

\begin{algorithm*}
\small
\caption{Sulfur emissions compliance smart contract.}
\label{alg:sulfur}
\begin{algorithmic}[1]
\Function{collectAndValidateSulfurEmissionsData}{}
    \State Initialize dataCollection object
    \State Initialize currentPosition $\gets$ \Call{getShipPosition}{}
    \State Initialize currentSulfurContent $\gets$ \Call{getSulfurContent}{}

    \Comment{Check if the ship is within an Emissions Control Area}
    \If{currentPosition.isInECA()}
        \State dataCollection.add(\textquotesingle Area\textquotesingle, \textquotesingle Emissions Control Area\textquotesingle)

        \Comment{Validate sulfur content for ECA}
        \If{currentSulfurContent $\leq$ 0.10}
            \State dataCollection.add(\textquotesingle Sulfur Content\textquotesingle, \textquotesingle Compliant - Below 0.10\%\textquotesingle)
        \Else
            \State dataCollection.add(\textquotesingle Sulfur Content\textquotesingle, \textquotesingle Non-compliant - Above 0.10\%\textquotesingle)
            \State \Call{commitToBlockchain}{currentPosition, currentSulfurContent}
        \EndIf

    \Else \Comment{Ship is outside of an Emissions Control Area}
        \State dataCollection.add(\textquotesingle Area\textquotesingle, \textquotesingle Non-ECA\textquotesingle)

        \Comment{Validate sulfur content for non-ECA}
        \If{currentSulfurContent $\leq$ 0.50}
            \State dataCollection.add(\textquotesingle Sulfur Content\textquotesingle, \textquotesingle Compliant - Below 0.50\%\textquotesingle)
        \Else
            \State dataCollection.add(\textquotesingle Sulfur Content\textquotesingle, \textquotesingle Non-compliant - Above 0.50\%\textquotesingle)
            \State \Call{commitToBlockchain}{currentPosition, currentSulfurContent}
        \EndIf
    \EndIf

    \State \Return dataCollection
\EndFunction
\end{algorithmic}
\end{algorithm*}

\subsubsection{Information flow and access}
The information flow within the system is managed through an access control list (ACL) and smart contracts, ensuring that only authorized entities can access the compliance data stored on the blockchain. The ACL specifies permissions for different stakeholders, including the vessel crew, ship owner, and flag state. The smart contracts enforce these permissions, providing secure and controlled access to compliance data. 

\begin{table}[ht]
\centering
\caption{Responsibility of each participant in the blockchain-enhanced compliance monitoring system.}
\resizebox{\columnwidth}{!}{%
\begin{tabular}{|l|c|c|c|c|}
\hline
\textbf{Participants' Responsibility} & \textbf{Vessel Owner} & \textbf{Ship's Crew} & \textbf{Flag State} & \textbf{Port State} \\
\hline
Install and maintain IoT devices/sensors & \checkmark & \checkmark & & \\
\hline
Deploy and manage smart contracts & \checkmark & & \checkmark &  \\
\hline
Upload compliance data to blockchain & \checkmark & & & \\
\hline
Validate blockchain data & \checkmark & & \checkmark & \\
\hline
View vessel compliance data & \checkmark & \checkmark & \checkmark & \\
\hline
Receive non-compliance notifications & & & \checkmark & \checkmark \\
\hline
\end{tabular}
}
\label{tab:participant_responsibility}
\end{table}

\section{Evaluation}  \label{sec:evaluation}

\subsection{Blockchain Network Setup and System Model}
The key components of the proposed model are the sensor data, polygon blockchain (Amoy testnet), and various system users.

\subsubsection{Sensor data}
Sensor data is collected from the vessel and uploaded directly onto the blockchain using satellite internet at fixed intervals. This data collection is facilitated by periodic data pulls from the sensors.

\subsubsection{Polygon blockchain}
The Polygon PoS blockchain functions as the distributed ledger, shared among all stakeholders in the peer-to-peer network. All transactions are recorded on this immutable ledger, ensuring data integrity and security. For development and testing purposes, the Amoy Testnet, which mirrors the Polygon Mainnet, is employed. This testnet provides developers with testnet tokens that hold no real value, allowing them to test system configurations and implementations without financial risk.

\subsubsection{System users}
The primary users of the proposed system include the vessel, vessel owner, vessel crew, flag state, and port state. Each user has specific roles and responsibilities in interacting with the blockchain network, contributing to the overall effectiveness and compliance monitoring of the maritime operations. The responsibilities of each stakeholder is described in Table \ref{tab:participant_responsibility}.

\subsection{Tools and Implementation}
Tools used for the implementation of the prototype include the following:

\subsubsection{Remix IDE}
Remix IDE is a user-friendly Integrated Development Environment for developing and deploying smart contracts in Solidity. It provides real-time code analysis, debugging tools, and a user interface for deploying contracts directly to the blockchain, streamlining the development process and ensuring robust smart contracts before deployment \citep{remixIDE}.

\subsubsection{MetaMask}
MetaMask is a cryptocurrency wallet that manages user accounts and digital assets, serving as a bridge between the browser and the Ethereum blockchain. It ensures all transactions, including sending tokens or deploying smart contracts, are confirmed by users, enhancing security and preventing unauthorized access \citep{metaMask}.

\subsubsection{Ganache}
Ganache provides a local blockchain environment on our CPU, simulating a blockchain network with ten dummy accounts preloaded with test Ether. This setup allows developers to test smart contracts and blockchain applications in a controlled environment, offering detailed insights for efficient debugging and optimization \citep{ganache}.

\subsubsection{Polygon blockchain (amoy testnet)}
The Polygon blockchain (Amoy testnet) offers a global test network that replicates the main Polygon network. It allows developers to experiment with applications using test tokens, providing a realistic environment to evaluate performance, scalability, and security before deployment on the live network \citep{polygonAmoy}.

\begin{figure}[h!]
  \centering
  \includegraphics[width=\linewidth]{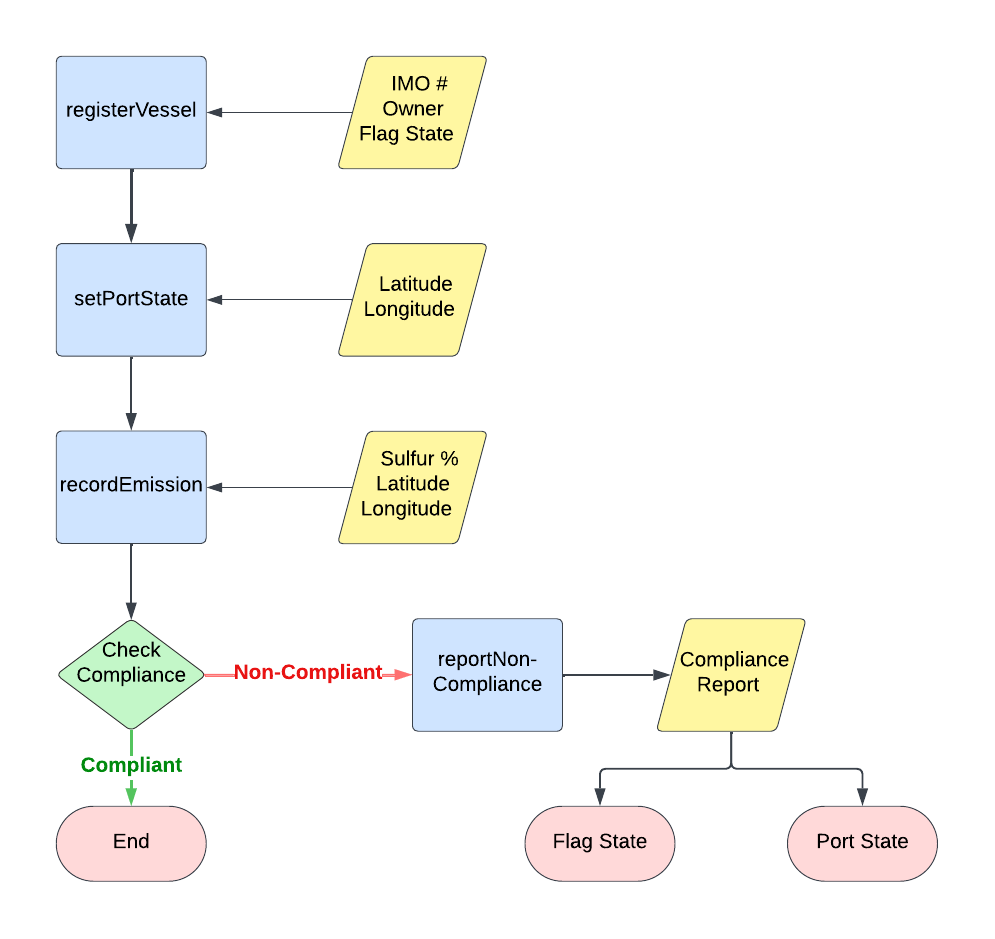} 
  \caption{Implemented smart contract flow.}
  \label{fig:contractflow}
\end{figure}

\subsection{Smart Contract Implementation}
The smart contract implementation for the proposed model is divided into three main Solidity files: 
\textit{VesselRegistration.sol}, \textit{Notification.sol}, and \textit{EmissionData.sol}. 
Figure~\ref{fig:contractflow} illustrates their interactions.

\subsubsection{VesselRegistration.sol}
The \textit{VesselRegistration} contract is responsible for managing vessel registration:
\begin{itemize}
    \item \textit{registerVessel}: Allows the admin to register a new vessel by providing its IMO number, owner, and flag state.
    \item \textit{isVesselRegistered}: Checks if a vessel is registered based on its IMO number.
    \item \textit{getFlagState}: Retrieves the flag state of a registered vessel using its IMO number.
\end{itemize}

\subsubsection{Notification.sol}
The \textit{Notification} contract manages non-compliance notifications and port states:
\begin{itemize}
    \item \textit{setPortState}: Allows the admin to set the port state for a given location (e.g., "USA").
    \item \textit{getPortState}: Retrieves the port state for a specified location.
    \item \textit{reportNonCompliance}: Records a non-compliance notification with details such as vessel ID, message, flag state, and port state. 
          This function is invoked automatically by the \textit{EmissionData} contract on non-compliance.
    \item \textit{getNotifications}: Returns an array of all recorded non-compliance notifications.
\end{itemize}

\subsubsection{EmissionData.sol}
The \textit{EmissionData} contract records sulfur emission data and checks for compliance:
\begin{itemize}
    \item \textit{recordEmission}: 
    Stores emission data for a vessel (sulfur content, position, ECA status). 
    The contract enforces regulatory limits (e.g., 0.10\% inside ECA, 0.50\% otherwise). 
    If emissions exceed the threshold, it automatically calls \textit{reportNonCompliance} in the \textit{Notification} contract.
    \item \textit{getEmissionHistory}: 
    Retrieves the full emission history for a given vessel, including timestamps, sulfur contents, ECA status, and compliance results.
\end{itemize}

\begin{figure*}[h!]
  \centering
  \includegraphics[width=\linewidth]{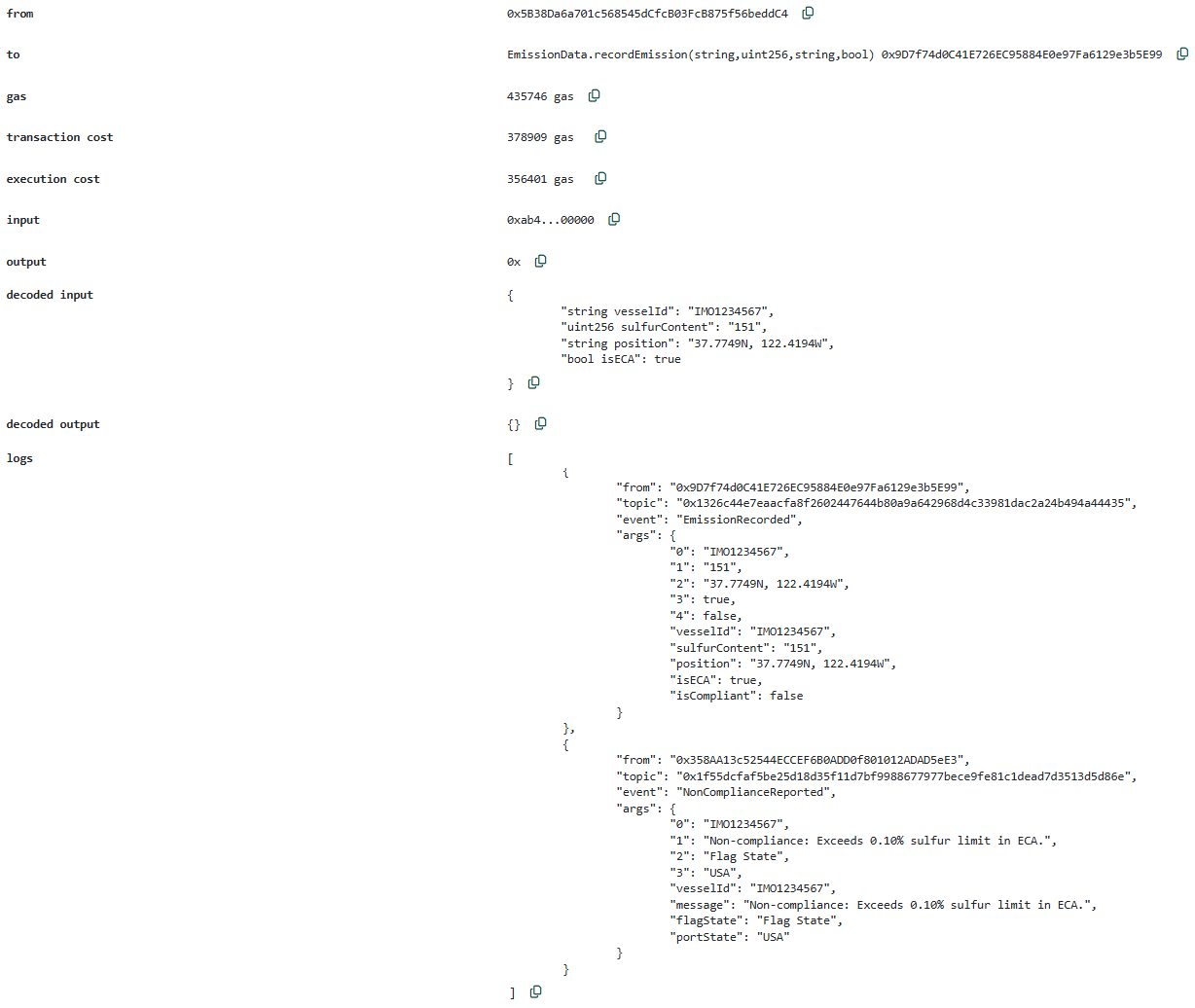} 
  \caption{Transaction details of executed recordEmission.}
  \label{fig:transactiondetails}
\end{figure*}

The transaction results of deploying and executing these contracts, with sample inputs, are illustrated in Figure~\ref{fig:transactiondetails}. 
The complete Solidity code is publicly available via GitHub.\footnote{\url{https://github.com/MarComplianceBlockchain/SystemPrototype}}

\subsection{Key Performance Metrics}
In evaluating the proposed blockchain-enhanced environmental compliance monitoring system for maritime operations, it is crucial to examine its scalability, throughput, operating cost, and security to ensure that the system can handle real-world operational demands efficiently. 

\subsubsection{Scalability}
Scalability refers to the system’s ability to handle increased loads without performance degradation. Our proposed system is hosted on the robust and efficient Polygon PoS network, a well-established public blockchain \citep{polygon}. This selection is crucial since the performance and scalability of our solution are inherently tied to the capabilities of the underlying blockchain. Polygon’s architecture allows for high throughput and low transaction costs while maintaining robust security measures. It is suitable for environments with frequent data transactions, such as maritime operations, where compliance data needs to be recorded and monitored at frequent intervals. It achieves this through an innovative combination of PoS and plasma sidechains. Unlike PoW, PoS does not require extensive computational resources, enabling the network to process more transactions per second. Sidechains further distribute the load and increase the capacity of the network. This architecture facilitates faster and more cost-efficient transactions than traditional blockchains like Ethereum. 

 \subsubsection{Throughput}
Throughput measures how many transactions a system can process in a given period. Throughput is critical for the proposed system as it directly impacts its ability to handle real-time compliance monitoring. If the system cannot keep up with the data flow, the blockchain structure allows for graceful degradation by temporarily queuing the incoming data. This results in delayed compliance notifications rather than data loss. Functionally, the impact of minor delays is minimal, as non-compliance reports would require further investigation by regulators before taking enforcement action.

\subsection{Scalability and Throughput Evaluation}

\begin{figure}[h!]
  \centering
  \includegraphics[width=\linewidth]{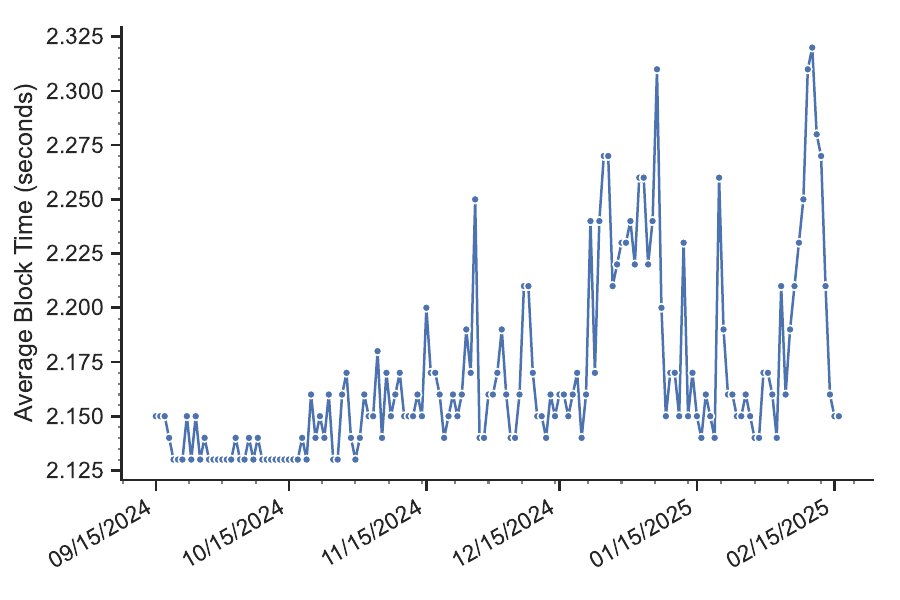} 
  \caption{Average block time for the last six months.}
  \label{fig:blocktime}
\end{figure}

\begin{figure}[h!]
  \centering
  \includegraphics[width=\linewidth]{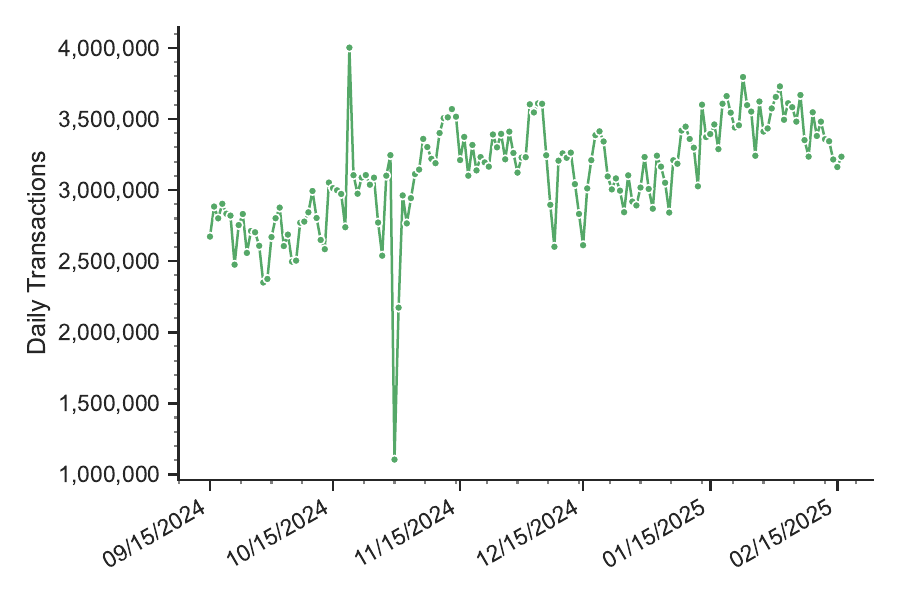} 
  \caption{Daily transactions for the last six months.}
  \label{fig:dailytransactions}
\end{figure}

\begin{figure}[h!]
  \centering
  \includegraphics[width=\linewidth]{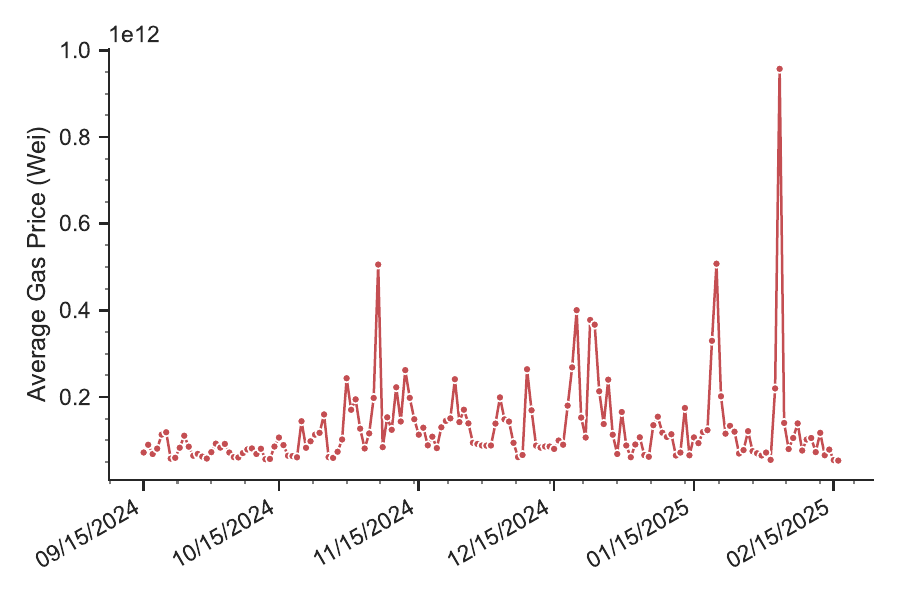} 
  \caption{Average gas cost for the last six months.}
  \label{fig:gascost}
\end{figure}

Evaluating the performance and scalability of a blockchain presents distinct challenges, such as the variability in transaction types, fluctuating gas fees, and the inherent lack of direct control over the network's nodes. Unlike traditional systems, where scalability is typically influenced by the computing power of the nodes, blockchain scalability is determined by the software managing block frequency and difficulty. To comprehensively assess potential performance, we analyze data on the number of daily transactions, average block time, and average gas fee prices on the Polygon blockchain over the past six months \citep{polygon}. Gas price, in this context, refers to the amount a user is willing to pay per unit of computational effort required to execute a transaction or smart contract. Based on the analysis of Figures \ref{fig:blocktime}, \ref{fig:dailytransactions}, and \ref{fig:gascost}, we derive several key conclusions.

\subsubsection{Average block time}
The average block time, which typically falls between 2.13 and 2.32 seconds, illustrates the network's capacity for scalability and consistent performance. This steadiness is beneficial for developers and users who depend on regular transaction and block processing times. Furthermore, the minor variations in block times suggest a resilient network infrastructure that sustains performance stability under diverse conditions, highlighting an efficiently optimized system.

\subsubsection{Daily transactions}
The fluctuation in daily transactions, ranging from approximately 1.4 million to 4 million transactions per day, demonstrates the Polygon network's ability to support high volumes of activity. The notable peaks in transaction volumes reflect periods of increased network use, indicating a growing and heavily utilized network, which is encouraging for the Polygon PoS blockchain adoption and usage.

\subsubsection{Average gas price}
The wide range in gas prices, from about 5.35e+10 to 9.57e+11 Wei, signifies a responsive and adaptable market within the Polygon network. This range indicates a robust ecosystem where network resources are priced based on demand. Additionally, the network's capability to handle significant spikes in gas prices shows its flexibility and scalability to manage surges in usage without becoming prohibitively costly for users.

\subsection{Gas Fee Evaluation}

\begin{table}[ht]
\centering
\caption{Gas cost analysis for smart contract deployment.}
\resizebox{\columnwidth}{!}{%
\begin{tabular}{|l|c|c|c|}
\hline
\textbf{Contract} & \textbf{Gas Cost} & \textbf{Gas Cost (MATIC)} & \textbf{Gas Cost (USD)} \\
\hline
VesselRegistration & 665,106 & 0.01995 & \$0.0066 \\
Notification & 1,188,150 & 0.03564 & \$0.0118 \\
EmissionData & 1,235,115 & 0.03705 & \$0.0122 \\
\hline
\end{tabular}
}
\label{table:deployment_gas_costs}
\end{table}

\begin{table}[ht]
\centering
\caption{Gas cost analysis for smart contract functions.}
\resizebox{\columnwidth}{!}{%
\begin{tabular}{|l|c|c|c|}
\hline
\textbf{Function} & \textbf{Gas Cost} & \textbf{Gas Cost (MATIC)} & \textbf{Gas Cost (USD)} \\
\hline
registerVessel & 95,741 & 0.0029 & \$0.0009 \\
setPortState & 49,077 & 0.0015 & \$0.0005 \\
recordEmission (Compliant) & 135,399 & 0.0041 & \$0.0014 \\
recordEmission (Non-Compliant) & 378,909 & 0.0114 & \$0.0038 \\
\hline
\end{tabular}
}
\label{table:gas_costs}
\end{table}

Tables \ref{table:deployment_gas_costs} and \ref{table:gas_costs} highlight the gas costs for the operations involved in the Maritime Compliance Blockchain framework. The deployment represents the costs of deploying each of the three smart contracts: VesselRegistration, Notification, and EmissionData on the Polygon Blockchain. These deployment operations are only carried out once per registered vessel during the initial setup of the system, with an approximate total cost of 0.0926 MATIC or \$0.31.

Additionally, the registerVessel function is utilized only at initial setup or if the owner of the flag of which the vessel is registered changes. The setPortState function and recordEmission functions are critical for monitoring sulfur emissions. These functions must both be called when recording emissions data, with setPortState followed by recordEmission. The setPortState function determines the vessel's position to identify which local authority, if any, would receive a non-compliant notification. The recordEmission function inputs the vessel ID, sulfur content from the sensor, and the vessel's position. It incurs different costs depending on whether the emission data is compliant or non-compliant. The cost of recording compliant emission data is lower than recording non-compliant data due to additional processing and notification steps involved in the latter. The overall cost of a compliant upload is \$0.003, and \$0.005 for a non-compliant upload. Assuming data pulls every hour, the approximate daily operating cost would be \$0.07 for compliant vessels. For the purposes of our cost analysis, expenses related to satellite connectivity are excluded, as nearly all MARPOL-compliant vessels are already equipped with satellite Wi-Fi systems.

One notable advantage of using the Polygon PoS blockchain is that reading data from the blockchain does not incur gas costs, which is beneficial for retrieving emission data and compliance notifications without additional expenses. The underlying network of the proposed framework demonstrates commendable performance stability, reflecting the reliability and efficiency of the Polygon Blockchain. Despite fluctuations in gas prices and transaction volumes, the network maintains consistent block times, showcasing its capacity to handle varying levels of activity and demand effectively. This analysis indicates that the Polygon PoS blockchain is a suitable platform for large-scale maritime compliance monitoring, providing a stable and cost-effective solution for real-time regulation enforcement. The network's performance and the efficiency of the smart contracts support the feasibility of deploying this framework in a real-world maritime environment.

\subsection{Security Evaluation} This section presents a security evaluation for the proposed blockchain-enhanced compliance monitoring system. We define a threat model and describing the system architecture, including measures to ensure data integrity before it reaches the blockchain, followed by a comprehensive security assessment of the blockchain.

\subsubsection{Threat Model and Adversary Capabilities} We consider two broad catagories of adversarial entities:
\textbf{(1) External attackers} targeting the network with spoofing, denial-of-service (DoS), or other exploits;
\textbf{(2) Internal actors} (e.g., ship owners, crew) who may tamper with sensors or attempt to submit false data to bypass regulations;

Adversaries may interfere at multiple stages: 
\begin{itemize} 
\item \textit{Sensor-Level Manipulation:} Disabling or spoofing sensors, altering recorded values at the source. 
\item \textit{Onboard Processing Interference:} Modifying validation logic or gateway software to inject false data. 
\item \textit{Network Attacks:} Eavesdropping, tampering, or DoS attempts during data transmission. \item 
\textit{Blockchain-Level Exploits:} Attempting 51\% or Sybil attacks to overwrite or fabricate ledger entries. \end{itemize}

\subsubsection{Network Architecture} Key procedures (\textit{CollectData}, \textit{ValidateData}) primarily run onboard the ship. Sensors and actuators communicate over industrial control system (ICS) protocols (e.g., Modbus, OPC UA) and are secured through firmware signing, role-based access, and tamper-evident hardware. Data is aggregated locally and transmitted via satellite or 5G to offsite nodes, where higher-level analytics may occur. Cryptographic keys are protected in hardware security modules, and regular patches minimize software vulnerabilities.

\subsubsection{Pre-Blockchain Data Integrity}
Since blockchain immutability alone cannot prevent fraudulent data at the source, we implement additional safeguards prior to data reaching the ledger:

\begin{itemize} \item \textit{Tamper-Resistant System Design:} Maritime environmental compliance systems require tamper resistent components, hardware, and software. These systems undergo rigorous governmental approval processes, reducing the likelihood of unauthorized modifications or data manipulation at the hardware level.

\item \textit{Calibration and Cross-Sensor Validation:} Periodic calibration under the oversight of maritime authorities prevents drift or deliberate misconfiguration. Redundant sensors, deployed at critical points, detect inconsistencies that could indicate physical tampering or sensor failures. Persistent deviations between sensor readings trigger alerts for further investigation by onboard staff or regulatory bodies.

\item \textit{Secure Transmission:} Shipboard data is encrypted in transit via TLS, while digitally signed logs ensure traceability of both sensor outputs and gateway actions. These logs are stored locally and can be audited offline to validate data integrity in the event of suspected system compromise.

\end{itemize}

\subsubsection{Blockchain Security Assessment}
Leveraging the Polygon blockchain, our proposed system integrates multiple security measures to ensure data confidentiality, integrity, and availability. Below is a concise summary of key threats and mitigations:

\begin{itemize}
\item \textit{Tampering}: Tampering involves unauthorized modifications of data to compromise integrity \citep{mohamed_rahouti_aa07229c}. The combination of cryptographic hashing and blockchain immutability detects and prevents retroactive alterations.

\item \textit{Information Disclosure}: Unauthorized access to sensitive data jeopardizes system confidentiality. In our system, end-to-end encryption and strict ACL policies limit data visibility to authenticated entities only.

\item \textit{Smart Contract Vulnerabilities}: Coding errors in smart contracts can inadvertently introduce security flaws \citep{zhang2019security}. We employ thorough contract audits and testing, which revealed no critical vulnerabilities in the current deployment.

\item \textit{51\% Attack}: In a 51\% attack, a single entity gains majority control of the network's computational power \citep{mohamed_rahouti_aa07229c}. Our approach uses a PoS consensus mechanism with ACL restrictions on validators, significantly reducing the likelihood of such an exploit.

\item \textit{Sybil Attack}: Sybil attacks entail creating numerous fake identities to influence the network. ACL-based identity verification and rate limiting prevent malicious actors from generating multiple illegitimate nodes.

\item \textit{Consensus Protocol Attacks}: Attacks on the underlying consensus algorithm aim to disrupt agreement on the network. Polygon PoS \citep{polygon} incorporates a robust, multi-layer consensus with frequent updates and failover measures, thereby deterring protocol-level exploits.
\end{itemize}

\section{Discussion and Conclusion}  \label{sec:discussion}
To the best of our knowledge, this study is among the first to explore how environmental compliance monitoring systems on ships, integrated with blockchain technology, can enhance regulatory adherence in real time. We presented a blockchain-enhanced framework that leverages the immutability and decentralization of a public ledger to secure critical compliance data, enabling robust real-time monitoring. Its decentralized nature also ensures that no single point of failure compromises the system, while smart contracts automate verification procedures and reduce the need for labor-intensive inspections.

A key advantage of blockchain in the global maritime domain is its suitability for multi-party collaboration among stakeholders with potentially competing interests (e.g., ship owners, flag states, and port states). Unlike a centralized or cloud-based solution, where a single entity controls the database, blockchain maintains a shared, tamper-resistant record that no party can retroactively alter. This setup offers an immutable audit trail of vessel data, encourages transparency by allowing stakeholders to independently verify records, and supports smart contract logic for near real-time enforcement of environmental rules. Such features are particularly relevant to the international, fragmented nature of shipping, where regulatory oversight spans multiple jurisdictions.

To demonstrate feasibility, we developed a proof-of-concept prototype using Polygon PoS for sulfur emissions monitoring. Experimental findings show that the model is both effective and efficient. Nonetheless, several practical challenges remain. Our current work is limited to a laboratory environment. Future research should involve extensive real-world trials, including additional shipboard systems for ballast water, oily waste, and sewage. These efforts should evaluate performance under a variety of operating conditions and involve larger networks of sensors and nodes.

Widespread adoption also depends on international regulatory support from the IMO and collaboration with industry stakeholders. For a fully operational system, sensors must be standardized and securely integrated with blockchain gateways that implement end-to-end encryption and access control. Notification mechanisms should be refined to ensure relevant authorities can promptly access compliance data. An alternative path could see individual nations deploy similar approaches for fleets under their jurisdiction. Further enhancements, such as artificial intelligence for anomaly detection and ongoing research on more scalable blockchain architectures, can further strengthen reliability and efficiency.

In conclusion, the proposed blockchain-enhanced compliance monitoring system offers significant promise for improving maritime environmental protection. By bringing together tamper-resistant data management, automated compliance checks, and transparent multi-stakeholder governance, our framework could substantially boost regulatory adherence, data integrity, and operational transparency, ultimately contributing to more sustainable global shipping practices.

\bibliographystyle{elsarticle-harv} 
\bibliography{references}

\end{document}